\documentclass[aps,prl,twocolumn,showpacs,amsmath,amssymb]{revtex4-1}
\usepackage{graphicx}
\usepackage{xspace}
\usepackage{color}

\newcommand{\FIG}[3]{
\begin{figure}[t]
\begin{center}
\includegraphics[width=\columnwidth]{#2}
\end{center}
\caption{#3}
\label{fig:#1}
\end{figure}}

\def\myo{myosin$\thinspace$II\xspace}
\def\Myo{Myosin$\thinspace$II\xspace}
\def\atp{ATP\xspace}
\def\adp{ADP\xspace}
\def\pho{P$_i$\xspace}

\def\kt{k_BT\xspace}
\def\nt{N_{\rm t}\xspace}
\def\nb{N_{\rm b}\xspace}
\def\fext{F_{\rm ext}\xspace}

\def\km{k_{\rm m}\xspace}
\def\kf{k_{\rm f}\xspace}
\def\vb{v_{\rm b}\xspace}
\def\vu{v_{\rm s}\xspace}
\def\veff{v_{\rm eff}\xspace}
\def\dw{d_{\rm w}\xspace}
\def\rd{\rho_{\rm d}\xspace}

\def\s{\operatorname{\rm s}}
\def\Hz{\s^{-1}}
\def\pN{\operatorname{\rm pN}}

\def\nm{\operatorname{\rm nm}}

\def\const{\operatorname{\rm const}}

\def\eq#1{Eq.\ \eqref{eq:#1}}

\def\fig#1{Fig.\ \ref{fig:#1}}

\newcommand{\Exp}[1]{\exp\left(#1\right)}
\newcommand{\Avg}[1]{\langle #1\rangle}
\newcommand{\dd}[2]{\frac{d}{d #2}#1}

\newcommand{\OLD}[1]{}

\begin{document}

\date{\today}

\title{Stochastic force generation by small ensembles of \myo motors}

\author{Thorsten Erdmann}
\affiliation{BioQuant, University of Heidelberg, Im Neuenheimer Feld 267, 69120 Heidelberg, Germany}
\affiliation{Institute for Theoretical Physics, University of Heidelberg, Philosophenweg 19, 69120 Heidelberg, Germany}

\author{Ulrich S.~Schwarz}
\email{Ulrich.Schwarz@bioquant.uni-heidelberg.de}
\affiliation{BioQuant, University of Heidelberg, Im Neuenheimer Feld 267, 69120 Heidelberg, Germany}
\affiliation{Institute for Theoretical Physics, University of Heidelberg, Philosophenweg 19, 69120 Heidelberg, Germany}

\begin{abstract}
Forces in the actin cytoskeleton are generated by small groups of
non-processive \myo motors for which stochastic effects are highly
relevant. Using a crossbridge model with the assumptions of fast
powerstroke kinetics and equal load sharing between equivalent states,
we derive a one-step master equation for the activity of a finite-sized
ensemble of mechanically coupled \myo motors. For constant external
load, this approach yields analytical results for duty ratio and
force-velocity relation as a function of ensemble size. We find that
stochastic effects cannot be neglected for ensemble sizes below $15$.
The one-step master equation can be used also for efficient computer
simulations with linear elastic external load and reveals the sequence of
build-up of force and ensemble rupture that is characteristic for
reconstituted actomyosin contractility.
\end{abstract}



\maketitle

Generation of motion and force by \atp-powered molecular motors is a
hallmark of living systems \cite{a:Howard1997}. In their cellular
environment, molecular motors usually operate in groups
\cite{guerin_coordination_2010}. A striking example is force generation
in skeletal muscle, where hundreds of non-processive \myo motors are
assembled into the thick filaments of the sarcomeres. Since the
pioneering work of Huxley \cite{a:Huxley1957}, the statistical physics
of large ensembles of \myo motors has been studied in great detail. It
has been shown that in order to describe the response of skeletal muscle
to varying loading conditions, it is essential that the unbinding rate
of \myo from actin is strain-dependent and decreases under load
\cite{a:Duke1999, a:VilfanDuke2003b}. In contrast to e.g.\ the
processive motor kinesin, this makes \myo a catch rather than a slip
bond \cite{veigel_load-dependent_2003, guo_mechanics_2006} and leads to
recruitment of additional crossbridges under load
\cite{a:PiazzesiEtAl2007}.

The collective activity of \myo motor ensembles is also essential for
the generation of motion and force in the actin cytoskeleton of
non-muscle cells. In this case, the actin structures are far more
disordered than in muscle and non-muscle \myo is usually organized in
mini-filaments comprising $10-30$ motors \cite{verkhovsky_myosin_1995}.
For such small numbers of motors, stochastic effects will become
important and are indeed observed in experiments. Measurements of
tension generated by \myo motors in reconstituted assays, e.g.\ in three
bead assays \cite{a:FinerEtAl1994, veigel_load-dependent_2003,
debold_slip_2005}, active gels \cite{a:MizunoEtAl2007,
silva_active_2011} or motility assays \cite{placais_spontaneous_2009},
reveal noisy trajectories, typically with a gradual increase of tension
followed by an abrupt release, which is likely due to detachment of the
whole ensemble (\textit{slip}). However, a detailed and analytically tractable
description for this biologically important situation is still missing.

The collective activity of mechanically coupled molecular motors has
been investigated before in the framework of a generic two-state
Fokker-Planck equation in which ensemble size enters into the noise
intensity \cite{a:JuelicherProst1997, placais_spontaneous_2009}. In
order to study effects of molecular details for ensembles of \myo
motors, crossbridge models originally developed for skeletal muscle can
be used as a starting point \cite{a:Duke1999, a:VilfanDuke2003b}. Due to
their complexity, these models are usually studied by computer
simulations. Analytical progress has been made with a mean field
approximation for large system size \cite{hexner_tug_2009}. Exploiting a
separation of time scales in the \myo cycle and using the assumption of
equal load sharing between motors in equivalent states, here we derive a
one-step master equation which explicitly includes the effects of catch
bonding and small system size. A one-step master equation has been
introduced before for transport by finite-sized ensembles of processive
motors with slip bond behavior \cite{a:KlumppLipowsky2005}, but not for
non-processive motors with catch bond behavior. Our results suggest that
stochastic effects are particularly important for ensemble sizes below
$15$, which corresponds to the typical size of cytoskeletal
mini-filaments.

\emph{Model.} We model the \myo cycle by three discrete mechano-chemical
states. The cycle is shown schematically in \fig{Cartoon}a. To allow for
comparison with earlier work, transition rates and most other molecular
parameters are taken from Refs.~\cite{a:Duke1999, a:VilfanDuke2003b}.
In practice, they will depend on \atp concentration and the exact type
of \myo \cite{veigel_load-dependent_2003,silva_active_2011}. In the
unbound state $(0)$, the motor-head is loaded with \adp and \pho and the
lever-arm is in its primed conformation. The motor then reversibly
transitions to the weakly bound state $(1)$ with forward rate $k_{01}
\simeq 40 \Hz$ and reverse rate $k_{10} \simeq 2 \Hz$. After release of
\pho, the lever-arm swings to the stretched conformation and the motor
enters the post-powerstroke state $(2)$. The transition rates between
the two bound states are relatively high, with $k_{12} \simeq k_{21}
\simeq 10^3 \Hz$. Replacing \adp by \atp, unbinding from the substrate
and hydrolysis of \atp brings the motor back to the unbound state $(0)$.
This last step is irreversible, with rate $k_{20} \simeq 80 \Hz$. Most
important in our context, both powerstroke and unbinding depend on load.
The powerstroke $(1) \to (2)$ moves the lever-arm forward by $d \simeq 8
\nm$ and strains the elastic neck-linker. Unbinding from $(2)$
requires further movement of the lever-arm, thus making unbinding slower under
load (\emph{catch bonding}).

\FIG{Cartoon}{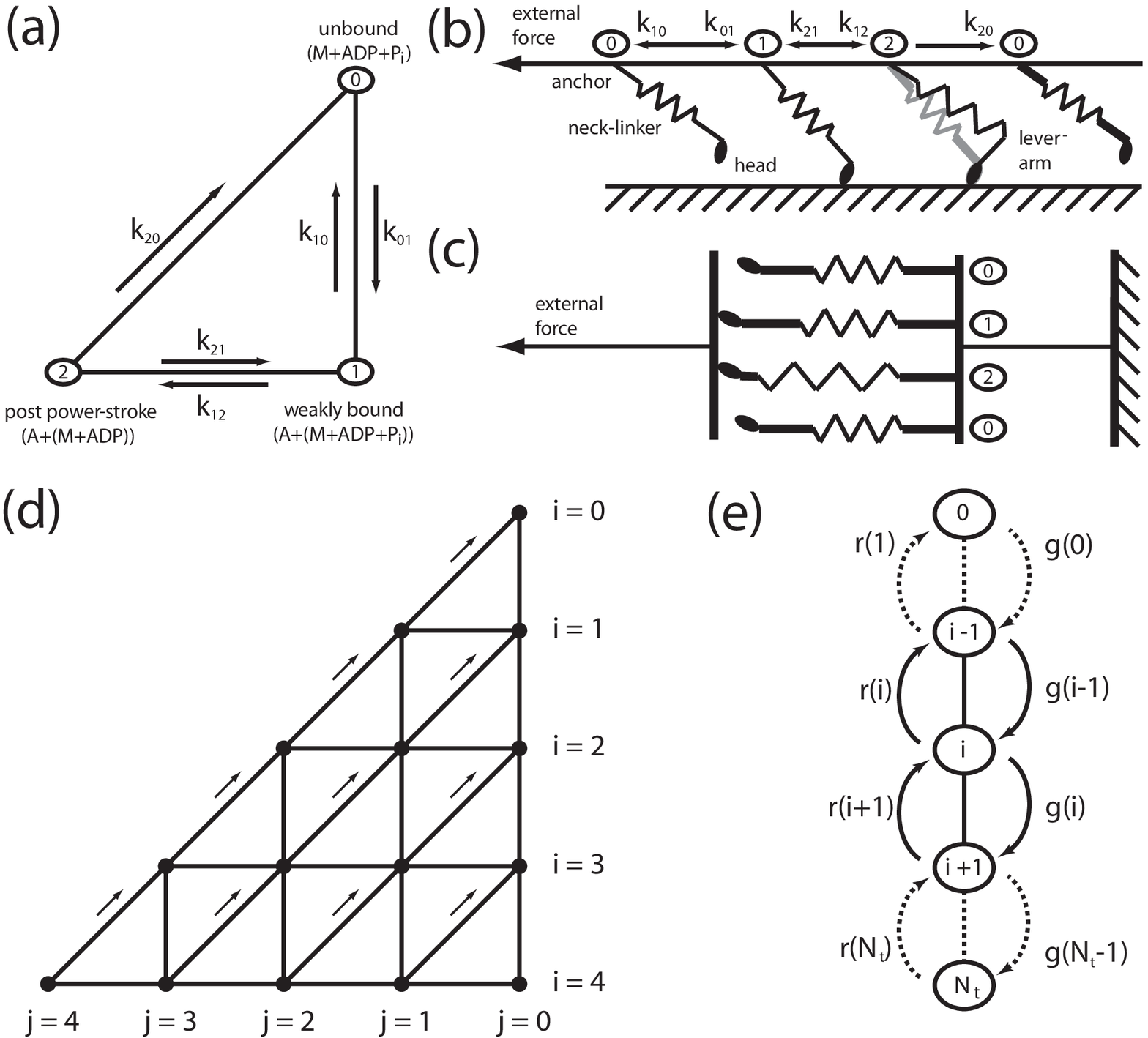}{(a) \Myo motor cycle with three mechano-chemical
states. (b) Mechanical coupling of a motor ensemble moving to the right.
The external load pulling to the left is balanced by elastic forces in
the neck-linkers of the motors. Swinging of the lever-arm increases the
strain of the neck-linker and hence the force exerted by a motor. (c) In
the \emph{parallel cluster model} (PCM), all motors in the same
mechano-chemical state have the same strain. (d) The corresponding
two-dimensional reaction network, with irreversible transitions marked
by arrows. (e) The effective one-dimensional network following from the
assumption of \emph{local thermal equilibrium} (LTE) of the bound
states.}

As shown schematically in \fig{Cartoon}b, the (upper) motor-filament
mechanically couples the different motors in an ensemble to each other.
Due to the strain-dependence of the rates, they are also dynamically
coupled. Hence, a complete description of the ensemble dynamics has to
include conformational state and strain of every motor. To arrive at a
tractable model, we first note that the motors pull in parallel. We next
assume that all motors in the same mechano-chemical state exert the same
force and hence have the same neck-linker strain. We thus arrive at the
\emph{parallel cluster model} (PCM) depicted in \fig{Cartoon}c, in which
the state of an ensemble with $\nt$ motors is characterized by the
number $i$ of bound motors and the number $j \le i$ of motors in the
post-powerstroke state. The number of motors in the weakly bound state
follows as $i-j$. In the PCM, each motor in the weakly bound state has the
same strain $x_{ij}$, where the indices indicate the dependence of the motor
strain on the ensemble state $(i,j)$. The powerstroke stretches the
elastic neck-linker by $d$, so that motors in the post-powerstroke
state have the strain $x_{ij}+d$. The strain
$x_{ij}$ of the weakly bound motors follows from the balance of the
external load $\fext$ and the elastic motor forces: $\fext = \km [(i-j)
x_{ij} + j (x_{ij} + d)]$. Here $\km \simeq 2.5 \pN \nm^{-1}$ is the
spring constant of the neck-linkers. For $\fext = \const$, the force
balance leads to
\begin{equation}\label{eq:Strain}
x_{ij} = \left(\fext - j \km d\right)/i \km\,.
\end{equation}
Thus the strain $x_{ij}$ of the weakly bound motors
is a state variable determined by external load
and both binding and powerstroke dynamics. If all motors are 
in the weakly bond state ($j=0$), it is positive. It can become negative
if sufficiently many motors have gone through the powerstroke 
and if the external load is not too large. The strain $x_{ij}+d$
of the post-powerstroke motors always stays positive and eventually
drives force generation and motion.

In the PCM, the network of reactions between states $(i,j)$ is
two-dimensional (see \fig{Cartoon}d). Due to slow binding and unbinding,
\emph{local thermal equilibrium} (LTE) is maintained for the bound
states \cite{a:VilfanDuke2003b}. When $i$ motors are bound, the
probability that $j$ motors are in the post-powerstroke state follows
the Boltzmann distribution $p(j|i) = \Exp{-E_{ij}/\kt} / Z$, where $Z$
is the appropriate partition sum. The energy $E_{ij} = E_{\rm el} +
jE_{\rm pp} + E_{\rm ext}$ in state $(i,j)$ is the sum of elastic energy
$E_{\rm el} = \km ((i-j) x_{ij}^2 + j (x_{ij} + d)^2)/2$ stored in the
neck-linkers, free energy bias $E_{\rm pp} \simeq -60 \pN \nm$ towards
the post-powerstroke state, and a possible external energy contribution
$E_{\rm ext}$. For $\fext = \const$, we have $E_{\rm ext} = 0$. 

LTE of the bound states allows us to project the $j$-axis onto the
$i$-axis, thus arriving at a one-dimensional reaction scheme with index
$i$ as shown in \fig{Cartoon}e. Then the probability $p_i(t)$ that $i$
motors are bound at time $t$ obeys the one-step master equation
\begin{equation}\label{eq:MasterEq}
\dd{p_i}{t} = r(i+1)p_{i+1} + g(i-1)p_{i-1} - [r(i)+g(i)]p_i\,.
\end{equation}
The probability to find an ensemble in state $(i,j)$ is $p_{ij}(t) =
p(j|i)p_i(t)$. The forward rate is $g(i) = (\nt-i)k_{01}$ because $\nt -
i$ free motors can bind.  Unbinding is possible from states $(1)$ and
$(2)$ so that the reverse rate for given $i$ and $j$ is $r(i,j) =
(i-j)k_{10} + j k_{20}(i,j)$. Averaging over $j$ gives $r(i) = \sum_j
p(j|i)r(i,j)$. The off-rate from state $(2)$ depends on the applied load
as $k_{20}(i,j) = k_0 \Exp{-\km (x_{ij} + d)/F_{0}}$, where $F_{0}
\simeq 12.6 \pN$. The strain-dependence of $k_{20}$ makes \myo a catch
bond. With these prescriptions, the one-step master equation
\eq{MasterEq} is fully specified for the case of constant external load,
$\fext = \const$. If the external load depends on the position of the
ensemble, like in the case of linear elastic loading, 
\eq{MasterEq} has to be solved together with additional prescriptions
for ensemble movement (see below).

\emph{Binding dynamics for constant load.} Mathematically, the reduction
to \eq{MasterEq} is a dramatic simplification, because many general
results are known for one-step master equations \cite{b:kamp92}. The
stationary distribution is
\begin{equation}\label{eq:StationaryDist}
p_i(\infty) = \frac{\prod_{j=0}^{i-1}\frac{g(j)}{r(j+1)}}{1 + \sum_{k=1}^{\nt}\prod_{j=0}^{k-1}\frac{g(j)}{r(j+1)}}\,.
\end{equation}
\fig{Binding}a plots the average number of bound motors, $\nb = \Avg{i}
= \sum_{i=0}^{\nt} i p_i(\infty)$, as function of $\fext$ for different
ensemble sizes $\nt$ (lines). The increase of $\nb$ is due to the catch
bond character of the post-powerstroke state. With increasing load,
$r(i,j)$ decreases to $(i-j)k_{10}$, so that $r(i)$ is small because
$p(j|i)$ is strongly biased towards large $j$ and $k_{10} \ll k_{20}$.
For skeletal muscle, the recruitment of additional crossbridges under
load has been observed experimentally \cite{a:PiazzesiEtAl2007} as
predicted by computer simulations \cite{a:Duke1999}. Here it follows in
a relatively simple way from analytical considerations. In order to
validate the PCM leading to \eq{MasterEq},
in \fig{Binding}a we also show results of computer
simulations which incorporate an individual strain value for each
motor (symbols). The agreement is very good, except at very small load, where
differences in the strain values between different motors
reduces unbinding, an effect which is less relevant under larger load.

\FIG{Binding}{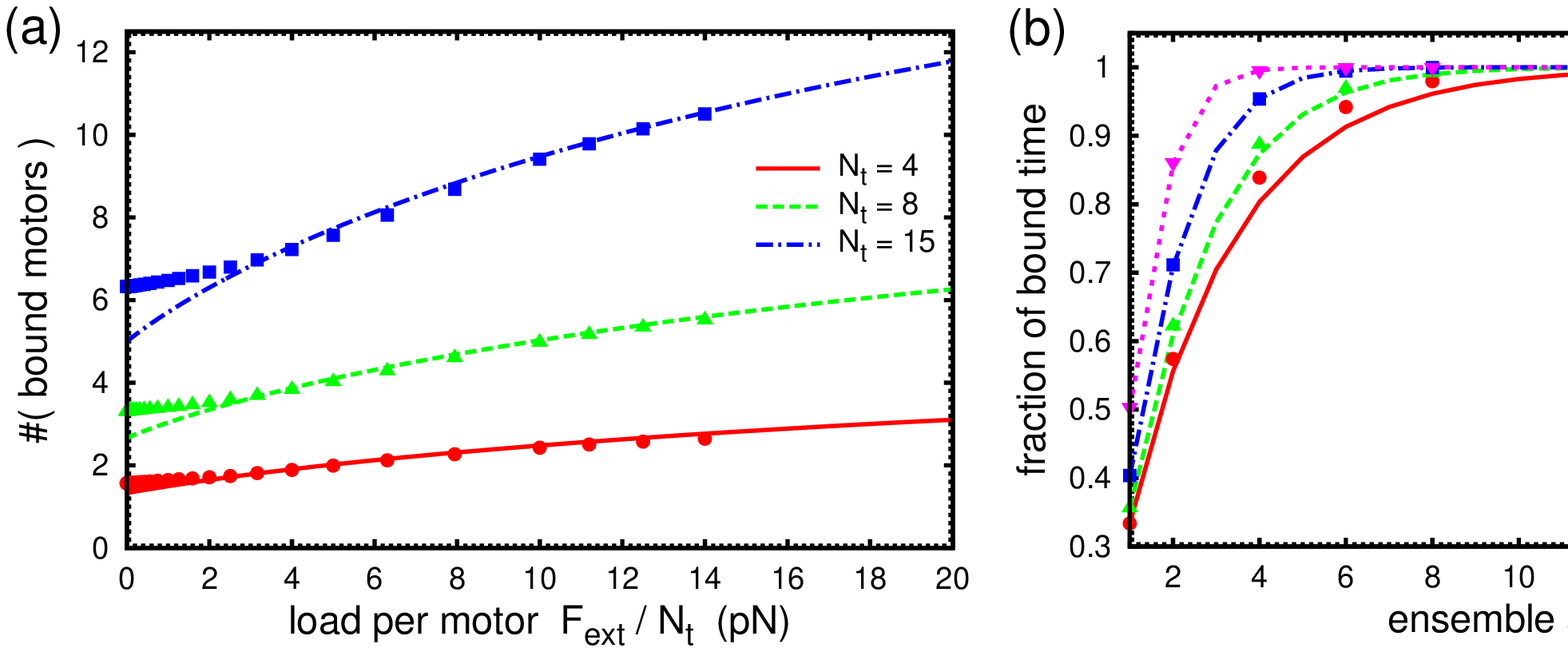}{(Color online) Analytical results for the parallel cluster
model (lines) and computer simulations with individual motor strains (symbols)
for constant external load. (a) Average number $\nb$ of bound motors following
from \eq{StationaryDist} as function of $\fext/\nt$ for $\nt = 4$, $8$
and $15$. (b) Duty ratio $\rd$ given by \eq{DutyRatio} as function of
$\nt$ for $\fext/\nt = 0.013 \pN$, $1.262 \pN$, $3.786 \pN$ and $8.834
\pN$.}

Next we discuss the effect of system size $\nt$. In general, smaller
ensembles are more likely to detach as a whole.
The mean first passage time for ensemble detachment after binding of
the first motor is
\begin{equation}
T_{10} = \sum_{j=1}^{\nt}\frac{1}{r(j)}\prod_{k=1}^{j-1}\frac{g(k)}{r(k)}\,.
\end{equation}
It is a polynomial of order $\nt-1$ in the ratio of binding to unbinding
rate and increases exponentially with ensemble size. Once the ensemble
has detached, on average it takes the time $T_{01} = 1/g(0) = 1/\nt
k_{01}$ to rebind.  We define the duty ratio of an ensemble as
\begin{equation}\label{eq:DutyRatio}
\rd = T_{10}/\left(T_{10} + T_{01}\right)\,.
\end{equation}
\fig{Binding}b plots $\rd$ as function of $\nt$ for different $\fext$ (lines).
Because $T_{10}$ increases and $T_{01}$ decreases with $\nt$, the duty
ratio increases quickly with $\nt$ and reaches unity for ensemble sizes
around $\nt \simeq 15$. With increasing force, $\rd$ increases faster
because of the increasing $\nb$. Again the agreement with
the simulation of the crossbridge model with individual motor strains
(symbols) is rather good except at very small force.
Stochastic effects are expected to be
important for duty ratios below unity, i.e.\ below ensemble sizes around
$15$. This implies that myosin mini-filaments in the cytoskeleton are
typically at the verge of stochastic instability.

\emph{Ensemble movement.} We now consider the spatial coordination
schematically depicted in \fig{Cartoon}b, that is, we assume an immobile
substrate over which an ensemble moves to the right. The PCM assumes
that all bound motor-heads are at the same position, which we denote by
the coordinate $z$. The anchors of the motors in the (upper)
motor-filament are located at the common position $z-x_{ij}$. Note that
whereas $z$ increases to the right in \fig{Cartoon}b, external load
$\fext$ and strain of the motors are defined in the opposite direction.
For the case of constant external load, $\fext = \const$, $z$ is a
variable which is slaved to the binding dynamics. When the ensemble
works against a linear external load, $\fext = \kf (z-x_{ij})$,
the value of $z$ enters the force balance and hence feeds back into the
system state. In addition, here one has to include an external elastic energy
$E_{\rm ext} = \kf (z-x_{ij})^2/2$, where $\kf$ is the external spring
constant.

Although more complicated assumptions might be possible, here we make
the following simple assumptions for the dynamics of $z$ within the PCM.
Starting with a state $(i,j)$, we assume that a motor binds to the substrate with vanishing strain at
the position $z-x_{ij}$ just below its anchor point. Binding of a
new motor thus changes the average position $z$ of the bound motor-heads
by $\Delta z = \left(iz + (z - x_{ij})\right)/(i+1) - z = - x_{ij} /
(i+1)$. Unbinding does not change $z$, because all bound motor-heads are
at the same position. The powerstroke does not change $z$ either,
because it does not affect the positions of the motor-heads. However,
$x_{ij}$ is affected by binding and unbinding as well as by the powerstroke
via the force balance, so that the position of the motor-filament
$z-x_{ij}$ is affected by all these transitions. When an ensemble
detaches completely from the substrate, the motor-heads
relax to the position $z-x_{ij}$ of the anchors. The detached ensemble
then moves backwards with velocity $\vu = -\eta \fext$ (\textit{slip}),
where $\eta$ is the effective mobility of the motor-filament. 

With these additional prescriptions, the rates defined for \eq{MasterEq}
can now be used to investigate the details of the stochastic movement of the motor ensemble
for arbitrary laws for the external load. To simulate stochastic
trajectories, we use the Gillespie algorithm \cite{a:Gillespie1976}.
After every change of $i$, $x_{ij}$ and $p(j|i)$ are updated to
calculate the average strain of the weakly bound motors,
$x_i=\sum_{j=0}^{i} x_{ij}p(j|i)$,
and the transition rates $r(i)$ and $g(i)$.
In case of binding, we change $z$ by $\Delta z = -x_i / (i+1)$.
\fig{Trajectory}a shows a stochastic trajectory of an ensemble
working against constant load. The lower panel shows the number of bound
motors $i$, the upper panel the average head position $z$ as function of
time. When bound, the ensemble moves forward with fluctuations around a
steady state velocity. A slip leads to backsteps of average size $\vu
T_{01}$. \fig{Trajectory}b shows a trajectory for an ensemble working
against a linear elastic load. The ensemble is slowed down by the load
building up by the forward motion. An increasing load stabilizes the
ensemble because $\nb$ increases. However, the very small ensemble
frequently detaches before reaching the stall force. Detachment leads to a
noisy trajectory in which the load fluctuates around an
effective stall force. This type of trajectories, with gradual buildup
and quick release of tension, resembles those experimentally observed in three bead
assays \cite{a:FinerEtAl1994, veigel_load-dependent_2003,
debold_slip_2005}, active gels \cite{a:MizunoEtAl2007,
silva_active_2011} and motility assays \cite{placais_spontaneous_2009}.

\FIG{Trajectory}{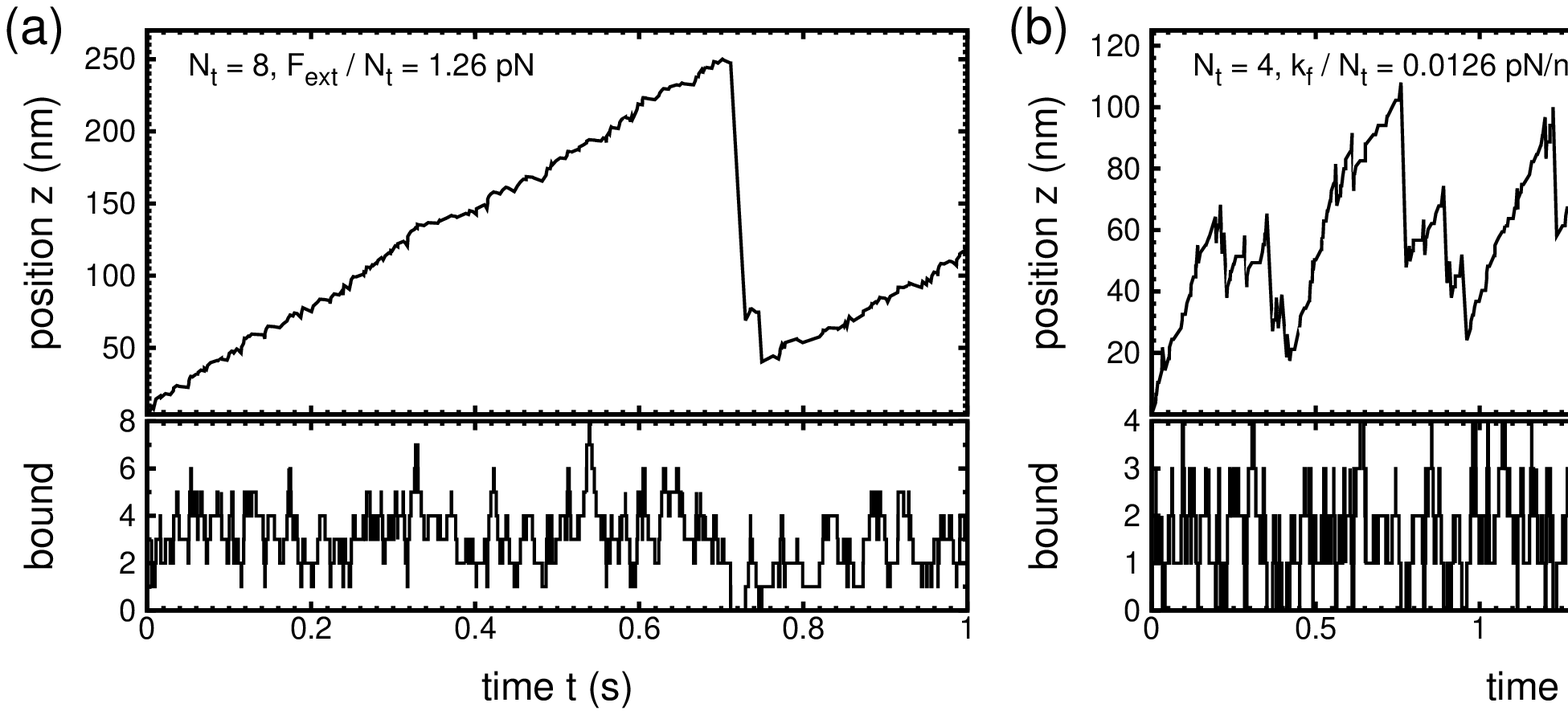}{Stochastic trajectories. (a) Constant load:
Average head position $z$ (upper panel) and number $i$ of bound motors (lower
panel) as function of time $t$ for ensemble size $\nt = 8$ and load
$\fext = 1.26 \pN \nt$. (b) Linear load: Average head position $z$ (upper panel) and
number $i$ of bound motors (lower panel) as function of $t$ ($\nt = 4$, $\kf /\nt =
0.0126 \pN\nm^{-1}$). In (a) and (b), an detached ensemble slides
backwards with mobility $\eta = 10^3 \nm \pN^{-1} \s^{-1}$.}

\emph{Force-velocity relation for constant load.} In state $(i,j)$, one
can identify the ensemble velocity with $v_{ij} = - g(i) x_{ij} / (i+1)$
(the ensemble only moves to the right when the strain
defined to the left is negative). The average stationary velocity
of a bound ensemble is
\begin{equation}\label{eq:BoundVelo}
\vb = \sum_{i=1}^{\nt} \sum_{j=0}^{i} v_{ij} p(j|i) p_i(\infty)
\end{equation}
with $p_i(\infty)$ from \eq{StationaryDist}. This is the force-velocity
relation of the bound ensemble at constant load. \fig{ForceVelocity}a
plots $\vb$ as function of the external load per motor for different
$\nt$. With increasing load, the velocity decreases. The upward convex
shape of $\vb(\fext)$ is due to the increase of $\nb$ with $\fext$,
which allows the ensemble to resist larger forces. For small ensembles for $\fext / N_t
> 0$, bound velocity $\vb$ and also the stall force increase with
increasing $\nt$. Above $\nt \simeq 15$, the force-velocity curve is
independent of $\nt$. This confirms our conclusion from the duty ratio
that stochastic effects cannot be neglected up to a system size of $15$
(compare \fig{Binding}b).

\FIG{ForceVelocity}{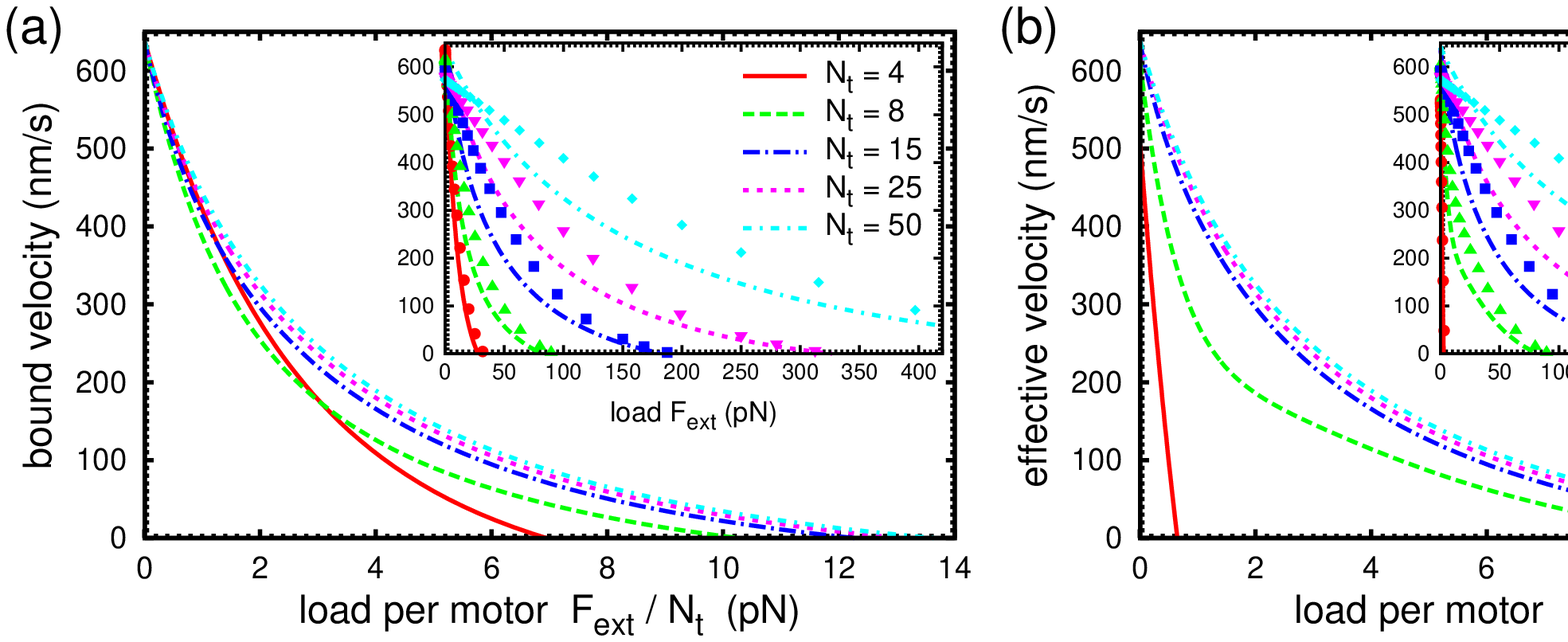}{(Color online) Force-velocity relation for constant load.
\emph{Main panels:} Analytical results for (a) the average bound
velocity $\vb$ (see \eq{BoundVelo}) and (b) the average effective
velocity $\veff$ (see \eq{EffectiveVelocity}) of an ensemble as function
of the external load per motor $\fext / \nt$ for $\nt = 4$, $8$, $15$,
$25$ and $50$. For $\veff$ the free mobility is $\eta = 10^3 \nm
\pN^{-1} \s^{-1}$. \emph{Insets:} Comparison of analytical results (lines)
with computer simulations of the crossbridge model 
with individual motor strains (symbols) for (a) $\vb$ and (b) $\veff$ as function
of external load $\fext$ for the same parameters as in the main
panels.}

Assuming that the stationary velocity is established quickly after
binding to a substrate, the walk-length of a motor-filament in one
attachment event is given by $\dw = \vb T_{10}$.
Although the bound velocity $\vb$ decreases, $\dw$ increases with
$\fext$ because the detachment time $T_{10}$ increases strongly. Only
upon passing the stall force, the walk-length drops to negative values.
Comparison with numerical solutions of the master equation (not shown)
reveals that $\dw = \vb T_{10}$ is a good approximation except for very
small values of $\dw$. Because the sliding velocity is negative, $\vu <
0$, the effective velocity is reduced by the occurrence of slip events:
\begin{equation}\label{eq:EffectiveVelocity}
\veff = \frac{\vb T_{10} + \vu T_{01}}{T_{10} + T_{01}} < \vb\ .
\end{equation}
\fig{ForceVelocity}b plots the effective velocity $\veff$ as function of
the external load per motor. Because the duty-ratio increases and the
rebinding time decreases with $\nt$, i.e., detachment is less frequent and
backsteps are smaller, the velocity at small $\fext$ now increases with
$\nt$. In addition, detachment of small ensembles leads to a faster
decrease of $\veff$ under load and a smaller stall force. Moreover,
detachment leads to large fluctuations of $z$ at the effective stall
force: instead of being stationary, the ensemble alternates between slow
forward motion when bound and fast backward slipping when detached. Above
the threshold of $\nt \simeq 15$, where the duty ratio is close to
unity, the effective velocity is identical to the bound velocity.

The insets in \fig{ForceVelocity} compare the analytical results using the
PCM to the computer simulations without PCM for $\vb$ and $\veff$ as
function of $\fext$. The agreement is rather good. Due to the molecular
friction resulting from differences in strain, the bound velocity at
vanishing load now decreases with $N_t$ and the curvature of the
force-velocity relation is less pronounced. The stall force and the
role of ensemble size for stochastic effects is predicted well.

\emph{Discussion.} In this Letter, we have derived a mathematically tractable model
for the collective behavior of small ensembles of \myo motors as a function of system size.
Our main assumption, the \emph{parallel
cluster model} (PCM) for the load sharing, was validated
by computer simulations of a cross-bridge model with individual
motor strains. These assumptions decrease
the disorder in the motor strains, so that the model cannot describe
powerstroke synchronization through load as it has been done before
with a detailed model for skeletal muscle \cite{a:Duke1999}. However, our model makes
accurate predictions for central quantities such as
duty ratio and force-velocity relation as a function of ensemble size.
For processive motors, the strains of the motors are homogenized
because fast moving motors are slowed down by the increasing
load. For the non-processive motors studied here, this mechanism
cannot operate. However, here the differences in the
strain of the bound motors are reduced by the small duty
ratio, thereby making the PCM a reasonable assumption for
our purposes. Due to its computational simplicity, in the
future the approach introduced here
can be used for studies of the intriguing interplay between actin filaments
and small ensembles of myosin II motors in the actin cytoskeleton of
non-muscle cells and reconstituted actomyosin systems. 

\begin{acknowledgments}
We thank Philipp Albert for helpful discussion. TE and USS were supported
by a Frontier-grant from Heidelberg University. USS is a member of
the Heidelberg cluster of excellence CellNetworks and was supported
through the MechanoSys-grant from the BMBF. 
\end{acknowledgments}


\end{document}